\documentclass{article}

\usepackage{arxiv}
\usepackage[table,xcdraw]{xcolor}
\usepackage{MnSymbol}
\usepackage{amsmath}
\usepackage{mathtools}
\usepackage{comment}
\usepackage{pgfplots}
\usepackage{biblatex}
\usepackage{pgf-pie} 
\usepackage[left]{lineno}
\usepackage[export]{adjustbox}
\usepackage{algorithm}
\usepackage{algpseudocode}
\usepackage{multirow}
\usepackage{tikz}
\usepackage{caption}
\usepackage{subcaption}
\usepackage{pgfplotstable}
\usetikzlibrary{trees,positioning,shapes,shadows,arrows}
\graphicspath{ {./images/} }
\usepackage{float}

\usepackage[utf8]{inputenc} 
\usepackage[T1]{fontenc}    
\usepackage{hyperref}       
\usepackage{url}            
\usepackage{booktabs}       
\usepackage{amsfonts}       
\usepackage{nicefrac}       
\usepackage{microtype}      
\usepackage{lipsum}		
\usepackage{graphicx}
\usepackage{doi}
\usepackage[export]{adjustbox}
\usepackage{multirow}
\usepackage{pgfplotstable}

\usepackage[edges]{forest}

\newtheorem{theorem}{Theorem}[section]

\newtheorem{definition}[theorem]{Definition}

\newif\ifadjust
\adjustfalse 

\title{Do not rug on me: Zero-dimensional Scam Detection}

\renewcommand{\shorttitle}{Do not rug on me:\, Zero-dimensional Scam Detection}

\hypersetup{
pdftitle={Do not rug on me:\, Zero-dimensional Scam Detection},
pdfsubject={q-bio.NC, q-bio.QM},
pdfauthor={Victor, Bruno, Vanesa,...},
pdfkeywords={Ethereum, DeFi, Dex, Scam Detection},
}
\author{
  \textbf{Bruno Mazorra}\\
  Nokia Bell-labs\\
  Universitat Pompeu Fabra\\
  \texttt{brunomazorra@gmail.com}\\
  \and
\textbf{Victor Adan}\\
  Universitat de Barcelona \\
  \texttt{victor8adan@gmail.com}
  \and
\textbf{Vanesa Daza}\\
  Universitat Pompeu Fabra \\
  \texttt{vanesa.daza@upf.edu}
} 

\begin{document}
\maketitle
\begin{abstract}
Uniswap, like other DEXs, has gained much attention this year because it is a non-custodial and publicly verifiable exchange that allows users to trade digital assets without trusted third parties. However, its simplicity and lack of regulation also makes it easy to execute initial coin offering scams by listing non-valuable tokens. This method of performing scams is known as rug pull, a phenomenon that already existed in traditional finance but has become more relevant in DeFi. Various projects such as \cite{rugpulldetector,tokensniffer}  have contributed to detecting rug pulls in EVM compatible chains. However, the first longitudinal and academic step to detecting and characterizing scam tokens on Uniswap was made in \cite{demystifying}. The authors collected all the transactions related to the Uniswap V2 exchange and proposed a machine learning algorithm to label tokens as scams. However, the algorithm is only valuable for detecting scams accurately after they have been executed. This paper increases their data set by 20K tokens and proposes a new methodology to label tokens as scams. After manually analyzing the data, we devised a theoretical classification of different malicious maneuvers in Uniswap protocol. We propose various machine-learning-based algorithms with new relevant features related to the token propagation and smart contract heuristics to detect potential rug pulls before they occur. In general, the models proposed achieved similar results. The best model obtained an accuracy of 0.9936, recall of 0.9540, and precision of 0.9838 in distinguishing  non-malicious tokens from scams prior to the malicious maneuver.
   
\end{abstract}
\section{Introduction}
Blockchain technology has proven to be enormously disruptive and empowering in both the public and private sectors of computing applications. Blockchains are permissionless and immutable digital ledgers that can be implemented and audited without a trusted third party or central authority. 
At their basic level, they enable participants to record transactions in a shared public ledger such that under the regular operation of the blockchain network, no transaction can be changed once committed. In 2008 \cite{nakamoto2019bitcoin}, the blockchain idea
was combined with several other technologies to create Bitcoin: a peer-to-peer electronic cash system protected through cryptographic mechanisms without needing a central repository or authority. However, users and developers perceived that Bitcoin had a limited use case due to the lack of complete programmability of the Bitcoin Virtual Machine. For this reason, many developers worked on the launch of other chains such as Ethereum, a Turing-complete  blockchain that has evolved to include a wide range of decentralized applications. 
Often, a decentralised application (DApp) will use one or more "smart contracts" deployed on top of the blockchain. \newline
A smart contract is an executable code that runs on the blockchain to facilitate, execute and enforce the terms of an agreement between untrusted parties. 
The most popular and exciting area where smart contracts have been crucial is decentralized finance (DeFi), which makes financial products available on a publicly decentralized blockchain network.  DeFi could potentially offer a new pseudo-anonymous, non-custodial and permissionless financial architecture that allows open audit \cite{werner2021sok}.
However, the Turing-complete blockchain technology is not a silver bullet; the pseudo-anonymous and permissionless nature of the blockchain allows attackers, scammers and money launderers to act with impunity.
In parallel to the work carried out in \cite{demystifying}, in this paper  we will focus on the thefts and scams in the most popular tool of DeFi, the Decentralized exchanges (DEX), the DeFi version of market exchanges. 
The most common way scammers and malicious agents execute a theft is through a rug pull.
A rug pull of a project is a malicious operation or set of operations in the cryptocurrency industry where the developers abandon the project and take the investors' funds as profits. As mentioned in \cite{demystifying}, rug pulls are a popular maneuver that usually happens in DEXs, particularly in Uniswap, where malicious agents develop an ERC-20 token\footnote{Ethereum Request for Comments 20} and list it on a DEX and pair it with a leading cryptocurrency like USD or Ether. 
Once some uninformed investors swap their leading coin for the token, the developers then remove all the currencies from the liquidity pool, making the token untradable and with zero economic value. 
In order to make the attack more profitable, the creators usually use different marketing tools such as a fake website, telegram groups, and Discord chat rooms to cultivate confidence among potential investors.
\newline

\textbf{Our contribution}: In this paper, we expand the rug pull dataset of the paper \cite{demystifying} to 27,588 tokens. To do this, we collected all Uniswap data until 03/09/2021 by directly interacting with the Ethereum blockchain. Then, we labelled different tokens as \emph{scam, malicious} and \emph{non-malicious} tokens using various relevant features of the smart contract and the liquidity pool state (see Sections \ref{subsection:token_propagation} and \ref{section:Class}). 
We manually observed different ways of executing the rug pull, proposed a rug pull classification, and observed new complex forms of performing the theft. Moreover, we have observed a further usage of the Uniswap protocol to send ETH \footnote{ETH is the native cryptocurrency of the Ethereum network} unnoticed for most common tracking protocols.
Finally, as we detail in Section \ref{section:Scam}, we propose new features of tokens, liquidity pools, and the transaction graphs; and a new framework to predict the probability of a liquidity pool becoming a rug pull or a scam in the future.\newline
In summary, in this work, we make the following contributions:
\begin{itemize}
    \item We provide the most extensive labelled dataset of Uniswap rug pulls to date, including the source code, the liquidity, the prices, the mint/burn, and transfer events. The dataset includes all tokens from $04/05/2020$  to $03/09/2021$. 
    In total, we labelled 26957 tokens as scams/rug pulls and 631 tokens as non-malicious.
    \item We provide a theoretical classification of three different types of rug pulls: simple, sell, and trap-door rug pulls, and provide tools to identify them.
    \item To the best of our knowledge, we are the first to design an accurate automated rug pull detection to predict future rug pulls and scams using relevant features of the pool's state   and the token distribution among the users. More specifically, we used the Herfindahl-Hirschman Index and clustering transaction coefficient as heuristics to measure the distribution of the token among the investors.
    \item We prove that the use of lock contracts such as Unicrypt by other scam detectors  \cite{tokensniffer,rugpulldetector} provided misleading data about the economic security of the token. More precisely, we show that 90\% of tokens using locking contracts tend to become a rug pull or a malicious token eventually.
    \item We define two methods that use Machine Learning models to discriminate between malicious and non-malicious tokens in different scenarios. In the first scenario, tokens can be evaluated at any block prior to the malicious maneuver. In the second scenario, all tokens are evaluated at a certain time after the creation of their respective pools.
    Specifically, we use a new Machine Learning algorithm based on attention mechanisms for tabular data called FT-Transformer\cite{FT-Transformer}. Our best model obtains an accuracy of 0.9936, recall of 0.9540 and precision of 0.9838 in distinguishing non-malicious tokens from scams in the first scenario and an accuracy of 0.992, recall of 0.784 and precision of 0.869 in the second scenario.

\end{itemize}
All these results can be replicated using the code and the pipeline in \cite{git}. To use it, we highly recommend access to a full or an archive Ethereum node.\newline

\textbf{Organization of the paper:} In Section \ref{section:related_work}, we describe the state of the art of scam detection in smart-contract-based blockchains. Section \ref{section:preliminars} gives an overview of DeFi and DEXs and the main features needed for our analysis. 
Section \ref{section:token_taxonomy} introduces a classification of malicious Uniswap maneuvers, emphasizing the theoretical methodology  behind different rug pulls. In Section \ref{section:data_section}, we explain the methodology used to obtain
all the data needed to train our models and obtain our results. In Section \ref{section:Class}, we explain the methodology used to label the tokens listed in the Uniswap protocol as malicious and non-malicious and give an overview of the results obtained by applying this methodology. Finally, Section \ref{section:Scam} explains the model used to detect future rug pulls in the early stages. We explain
the two different methodologies used and describe the accuracy, sensitivity, and F1-score of the models used.
\newpage
\section{Related work}\label{section:related_work}

Due to the role of smart contracts in Blockchain Technology, some studies have analyzed the security and the automatic vulnerability detection of within such contracts.
Some have focused on finding anomalies in the transaction graph \cite{li2020dissecting,oforiboateng2021topological,2005,patel2020graph} and clustering malicious addresses \cite{9169433,9223304,seres2019mixeth}. 
For example, paper \cite{li2020dissecting} uses the transaction graph to predict relevant market price changes and  \cite{seres2019mixeth} uses the transaction graph and fingerprints in the gas price\footnote{https://ethereum.org/en/developers/docs/gas/} in order to detect the addresses of the same user. \newline
Other studies have focused on the vulnerabilities of smart contracts  \cite{dika2018security,feist2019slither,oliva2020exploratory,torres2019art}, for example \cite{feist2019slither} is a static analysis framework for smart contracts that detects potential vulnerabilities.
In a similar direction, other research \cite{babel2021clockwork,caldarelli2021blockchain,flashboys,eskandari2021sok,highfrequency} digs into vulnerabilities of DeFi protocols when interacting with rational agents.\newline
Regarding blockchain scams, many studies have investigated phishing scams \cite{ijcai2020-621,9184813,9493255,9180815}, Ponzi Schemes \cite{bartoletti2020dissecting,8668768,liebau2019crypto}, and automated scam detection \cite{9229769,9354770,9145151}. \newline 
More related to our work, various studies or projects \cite{rugpulldetector,tokensniffer,demystifying} have addressed the detection of rug pulls or frauds working on top of DEX protocols. Two projects \cite{rugpulldetector,tokensniffer} use the simple heuristics of holders, liquidity and an automatic smart contract analysis, to give a risk score of the token. Both projects share two major problems: the lack of longitudinal studies to check their results, and the failure to detect non-malicious tokens accurately. 
On the other hand, \cite{demystifying} provided the only longitudinal and cross-sectional study to date. The study provides a good introduction and overview to different rug pulls, as well as a dataset of more than 10K scam tokens listed in Uniswap. However, the major flaw of that paper is that the algorithm is trained to classify tokens and detect rug pulls only after they have occurred; that is the machine learning algorithm is trained in order to classify tokens, but is not able to detect future rug pulls.

\section{Preliminaries}\label{section:preliminars}
\subsection{Ethereum and Smart Contracts}\label{subsection:ethereum}
Ethereum is a blockchain with a quasi-Turing complete programming virtual machine, that compiles different programming languages such as Solidity\cite{dannen2017introducing}. One relevant goal of Ethereum is to any party to develop arbitrary applications and scripts that
execute in blockchain through transactions, using the blockchain to synchronize their state in a manner that is fully verifiable by any system participant. These scripts are usually refereed as smart contracts. Participants and smart contracts
in the Ethereum network transact with the base currency known as Ether. Ether, is the coin used to transact and to pay the fees to  the miners to transfer Ether or to interact with smart contracts. 
Accounts on the Ethereum network can be linked to programs in a virtual machine-based language called the Ethereum Virtual Machine (EVM).
More specifically, smart  contracts  are  programs  which  are  deployed  on  the  blockchain public ledger and are executed in transactions, which, similarly to ACID-style database transactions \cite{babel2021clockwork}, alter the state of the ledger atomically (that is, either all the operations of the transaction are executed or all the operations are reverted). In the moment of deploying a smart contract, a byte-code is sent in a transaction to the ledger, and this contract is assigned a unique address of 42 hexadecimal characters and its code is uploaded to the ledger. Once successfully  created,  a  smart  contract  consists  of  a  contract address of 42 hexadecimal characters, a balance, a code defined in the contract creation,  and  a  state. Different users and parties can then change the state of a  specific  contract by sending transactions invoking particular functions to a known contract address. If the transaction hold the constraints hard-coded in the smart contract, this transaction will trigger a set of actions established in the smart contract code as a result, such as reading and modifying the contract state, interacting and executing other contracts, or transferring Ether or tokens to other addresses. These actions can be coded to produce events, a transaction log of the relevant information produced by the actions triggered. These events are useful for developers and users to track the state of the smart contract. \newline
 The most popular and significant smart contracts of Ethereum are known as ERC-20 tokens, and emerged in 2015 as the technical standard used for all smart contracts on the Ethereum blockchain for fungible token implementations. A token is fungible if any token is exactly equal to any other token; no tokens have special rights or behavior associated with them. This makes ERC-20 tokens useful for currency exchange, voting rights, staking, and more. 
ERC-20 defines a common set of functions, of which only the signatures, but not the implementations, are specified. The table \ref{table:erc20} lists the common rules of ERC-20, including the global variables and functions.

\begin{table}[!h]
\centering
\begin{tabular}{|lll|}
\multicolumn{3}{l}{}                                                                                                                                                                         \\ \hline
\multicolumn{1}{|l|}{{\color[HTML]{000000} Requirements}}                & \multicolumn{1}{l|}{{\color[HTML]{000000} Type}}                     & {\color[HTML]{000000} Signature}                                        \\ \hline
\multicolumn{1}{|l|}{{\color[HTML]{000000} }}                           & \multicolumn{1}{l|}{{\color[HTML]{000000} }}                         & {\color[HTML]{000000} $\texttt{totalSupply()}$}                         \\ \cline{3-3} 
\multicolumn{1}{|l|}{{\color[HTML]{000000} }}                           & \multicolumn{1}{l|}{{\color[HTML]{000000} }}                         & {\color[HTML]{000000} $\texttt{balanceOf(address)}$}                    \\ \cline{3-3} 
\multicolumn{1}{|l|}{{\color[HTML]{000000} }}                           & \multicolumn{1}{l|}{{\color[HTML]{000000} }}                         & {\color[HTML]{000000} $\texttt{transfer(address,uint256)}$}             \\ \cline{3-3} 
\multicolumn{1}{|l|}{{\color[HTML]{000000} }}                           & \multicolumn{1}{l|}{{\color[HTML]{000000} }}                         & {\color[HTML]{000000} $\texttt{approve(address,uint256)}$}              \\ \cline{3-3} 
\multicolumn{1}{|l|}{{\color[HTML]{000000} }}                           & \multicolumn{1}{l|}{{\color[HTML]{000000} }}                         & {\color[HTML]{000000} $\texttt{allowance(address,address)}$}            \\ \cline{3-3} 
\multicolumn{1}{|l|}{{\color[HTML]{000000} }}                           & \multicolumn{1}{l|}{\multirow{-6}{*}{{\color[HTML]{000000} Method}}} & {\color[HTML]{000000} $\texttt{transferFrom(address,address,uint256)}$} \\ \cline{2-3} 
\multicolumn{1}{|l|}{{\color[HTML]{000000} }}                           & \multicolumn{1}{l|}{{\color[HTML]{000000} Event}}                    & {\color[HTML]{000000} $\texttt{Transfer(address,address,uint256)}$}     \\ \cline{3-3} 
\multicolumn{1}{|l|}{\multirow{-8}{*}{{\color[HTML]{000000} Required}}} & \multicolumn{1}{l|}{{\color[HTML]{000000} }}                         & {\color[HTML]{000000} $\texttt{Approval(address,address,uint256)}$}     \\ \hline
\multicolumn{1}{|l|}{{\color[HTML]{000000} }}                           & \multicolumn{1}{l|}{{\color[HTML]{000000} }}                         & {\color[HTML]{000000} $\texttt{name()}$}                                \\ \cline{3-3} 
\multicolumn{1}{|l|}{\multirow{-2}{*}{{\color[HTML]{000000} Optional}}} & \multicolumn{1}{l|}{{\color[HTML]{000000} }}                         & {\color[HTML]{000000} $\texttt{symbol()}$}                              \\ \cline{3-3} 
\multicolumn{1}{|l|}{{\color[HTML]{000000} }}                           & \multicolumn{1}{l|}{\multirow{-3}{*}{{\color[HTML]{000000} Method}}} & {\color[HTML]{000000} $\texttt{decimals()}$}                            \\ \hline
\end{tabular}
\caption{ERC-20 standard signatures.}
\label{table:erc20}
\end{table}

In addition to the base ERC-20 functionality, many tokens provide other functionalities \cite{oliva2020exploratory}. For instance, it is quite common to find contracts that can freeze accounts, transfer ownership, pause contracts or make complex interactions with other DeFi protocols. In this study, we focused on three functionalities that involve manipulation of tokens: minting, pausable, and complex buy/sell operations in which tokens can be obtained from or exchanged to Ether. Token minting corresponds to the creation of tokens, increasing the total supply of tokens and associating the newly minted tokens to a specific address. Token burning is the reverse operation: tokens can be erased from an account and their total supply decreases. Token sale works in terms of operations that allow an account to buy tokens using Ether, or obtain Ether by selling tokens.\newline
In order to obtain those features, one can use compilers such as Slither \cite{feist2019slither}, a static analysis framework designed to provide human-readable information and insights of smart contracts wrote in Solidity programming language. Slither allows the application of commonly used program analysis techniques like dataflow and taint tracking. Moreover, Slither detects various important features and vulnerabilities, like minting, reentrancy vulnerability, and pausable smart contracts.

\subsection{Decentralized Exchanges}
Decentralized Exchanges (DEXs) \cite{werner2021sok} are a category of Decentralized Finance (DeFi) protocol that allow the non-custodial exchange of digital assets. All trades are executed on-chain and are, thus, publicly verifiable. The policy that matches buyers and  sellers (or traders and liquidity providers) is hard-coded in a smart contract. DEXs have different mechanisms for price discovery: order book DEXs and automated market makers (AMM).
While order book exchanges have been broadly studied \cite{puljiz2018market,warren20170x} and used in traditional finance (TradFi), AMMs have been proven to be more useful in blockchain environments due to their computational efficiency and simplicity \cite{werner2021sok}.
In general, in an automated market maker, each asset pair comprises a distinct pool or market. Liquidity providers supply liquidity by adding both assets in proportion to the existing pool size. Traders exchange assets by adding one asset to the pool and removing the other. The ratio of the two traded assets is the average price paid, is calculated according to a downward sloping, convex relationship called \emph{constant function} (CF). The convexity implies that the AMM is liquidity sensitive, that is, larger orders have a larger price impact.
 DEXs and, in particular AMM, have become very popular in DeFi for several reasons: 
 \begin{enumerate}
     \item they permit easy provision of liquidity for minor assets, that is, any assets can be listed in a DEX,
     \item they allow any party to become a market maker,
     \item they are censorship-resilient in highly volatile periods,
     \item they can be audited by anyone.
 \end{enumerate}
 However, these properties also have their drawbacks: 
 \begin{enumerate}
     \item blockchain transactions are publicly visible in the mempool, which means that  miners or users can front-run trading transactions in DEXs \cite{flashboys,highfrequency}, consistently leading to a worse price for users,
     \item every token can be listed to trade in AMM protocols making uninformed users to fall into different scams or suboptimal performing projects \cite{demystifying}.
 \end{enumerate}

With an AMM, the price of an asset is deterministic by the state (number of reserves) and the number of assets that users are willing to trade. The most popular AMMs such as Uniswap\footnote{https://uniswap.org/whitepaper.pdf}, Sushiswap\footnote{https://sushi.com/}, Curve \footnote{https://curve.fi/whitepaper} and Balancer \footnote{https://balancer.fi/whitepaper.pdf} use different variations of the constant product formula, see Section \ref{Uniswap}.
\newpage
\subsection{Uniswap}\label{Uniswap}
Uniswap, the most relevant decentralized exchange, was launched in November 2018 and, to date, more than $40,000$ ERC-20 tokens are locked and tradable in the Uniswap protocol, adding a total value of 7 billion USD. In this section, we will provide an overview of the Uniswap V2 protocol for more details see \cite{adamsuniswap,angeris2019analysis}.
\newline
\textbf{Providing Liquidity}: Each pair pool comprises a pair of tokens. Most frequently,
as we will show in Section \ref{section:data_section}, one of the currencies is wrapped Eth (Weth), the ERC-20 equivalent version of Ether. We will typically use Eth or Weth as the numéraire and will denote it by $\mathcal E$, and we will refer to the other ERC-20 tokens as token, denoted by $\mathcal C$. A party wishing to provide liquidity to a specific pool deposits both  $\mathcal E$ and
 $\mathcal C$ into the pool. If the pair pool has no tokens deposited yet, the deposit ratio can be arbitrarily chosen by the liquidity provider. Otherwise, the deposit ratio of Eth to token is determined by the existing ratio in
the pool, which implicitly defines the infinitesimal price of the token $\mathcal C$ with respect to Eth $\mathcal E$.
A liquidity provider who makes such a deposit receives a proportional amount of a liquidity provider token (LP-token).
This third token is specific for each pool listed in Uniswap and represents the share of the liquidity provided by the agent of the total liquidity pool.
As the users swap tokens in the pool, the value of the liquidity pool may rise or fall in value. 
Liquidity providers can redeem their liquidity tokens at any time and get their share of the liquidity pool paid out in equal value of Eth and tokens. 
Providing liquidity is potentially
profitable because each trade incurs a transaction fee of $0.3\%$ which is redeposited into the pool. However, providing liquidity also has its own risks, leading in some situations to temporary losses \cite{aigner2021uniswap}.
\newline
\textbf{Price formula}: The pricing protocol for tokens listed on Uniswap is given by a constant product formula \cite{angeris2019analysis}. Suppose that a trader wants to buy an amount of $\Delta_y$ tokens, and the current reserves of the pair pool of Eth and tokens are $x$ and $y$ respectively. Then the amount $\Delta_x$ of Eth that they have  to deposit is the unique solution to the following equation
\begin{linenomath*}
\begin{equation*}
    (x+(1-f)\Delta_x)(y-\Delta_y) = xy,
\end{equation*}
\end{linenomath*}
where $f$ is the fee of the protocol, currently sett to $f=0.3\%$. After the trade, the reserves of the pair pool are updated in the following way $x\leftarrow x+\Delta_x$, $y\leftarrow y-\Delta_y$. Analogously, if a trader wants to sell tokens for Eth. 
The name \emph{constant product market} comes from the fact that when the fee is set to zero, any trade must change in a way that the reserves lies in the curve $xy = k$ for some positive real number $k$.
\newline
\textbf{Uniswap Architecture}:
Uniswap V2 contracts are divided into two types of contracts, the \emph{core} and the \emph{periphery}. This division allows the core contracts, which hold the assets and therefore have to be secure, to be audited more easily. All the extra functionality required by traders can then be provided by periphery contracts.
The most relevant periphery contract is the UniswapV2Router\footnote{https://etherscan.io/address/0x7a250d5630b4cf539739df2c5dacb4c659f2488d}. This contract allows the user to easily interact with other core contracts in order to quote prices, create pair pools, swap tokens and add/remove liquidity.
Two of the most fundamental core contracts are the UniswapV2Factory\footnote{https://etherscan.io/address/0x5c69bee701ef814a2b6a3edd4b1652cb9cc5aa6f} and UniswapV2Pair\footnote{https://github.com/Uniswap/v2-core/blob/master/contracts/UniswapV2Pair.sol}. The UniswapV2Factory is responsible for creating new pool pairs and recording all the pairs created. The UniswapV2Pair contract is responsible for recording the current state of the pool, i.e. balance of Eth $\mathcal E$ (or any other token $\mathcal C'$) and the token $\mathcal C$, computing the price for trading and the number of tokens needed to add liquidity. Moreover, the UniswapV2Pair contract has an ERC-20 structure and records the ownership of the liquidity provided to the pool.\newline
\textbf{Uniswap Events}: As we mentioned in \ref{subsection:ethereum}, events track the state of different variables of a smart contract. The Uniswap protocol contains five important events.
\begin{enumerate}
    \item \texttt{PairCreated}: It is an event in the UniswapV2 Factory contract. This event  emits each time a new pair is created, and outputs the tuple (\texttt{token0,token1},\texttt{pair},\newline
    \texttt{block\_creation}) of a new pool created.
    \item \texttt{Sync}: It is an event in the UniswapV2 PairPool contract. This event emits each time the reserves of the pool change. Every time the balance of the pool updates, the smart contract outputs the tuple (\texttt{reserves0,reserves1}), that is the reserves of the \texttt{token0} and \texttt{token1} after the update.
    \item \texttt{Mint,Burn \& Transfer}: These are events in the UniswapV2 PairPool contract that tracks the state of the ERC-20 LP-token. 
\end{enumerate}
\textbf{Locking contracts}: The locking contracts are protocols that run on top of the Uniswap protocol to provide a partial solution to rug pulls. These protocols are not part of the main Uniswap protocol. In general, in the first phase of developing a token or project, the liquidity provided to Uniswap is mostly added by the developers or creators of the project. It is for this reason that initially,  the distribution of the LP tokens is managed by a small number of addresses, making potential investors less confident in  the project. In order to provide some trust to new investors, the developers lock the liquidity in a smart contract (Unicrypt is the most popular DEX LP lock) or burn the LP token, making it unfeasible for the developer to remove the liquidity for some time or indefinitely.

\subsection{Token Propagation}\label{subsection:token_propagation}
We refer to Token propagation to the set of metrics and tools to study the token distribution and circulation of the token during its activity period.
As we mentioned previously, tokens are transferable. To send a token from an address $A$ to an address $B$, the sender $A$ can either call the function $\texttt{transfer}$ of the token's smart contract or call other smart contracts that inherit the functions $\texttt{approve}$ and $\texttt{transferFrom}$. Either way, an ERC-20 token emits the event $\texttt{transfer}$ that contains the tuple $(\texttt{sender,receiver,amount})$. 
The first way of transferring tokens is usually cheaper and is used to directly transfer the tokens from one external ownable account (EOA) to another. The most common approach is to deposit or withdraw tokens from centralized exchanges. The second way tends to be used by different DeFi protocols such as DEXs, lending protocols, or voting systems to allow smart contracts to exchange tokens on behalf of an EOA address. 
\subsubsection{Token distribution}
The set of transfers and transactions allows us to compute the balance of each address for any snapshot. In order to study the distribution of the tokens, we propose using the Herfindahl-Hirschman Index (HHI). \newline
In a nutshell, the HHI is a popular measure of market concentration and is used to calculate market competitiveness. The closer a market is to a monopoly, the higher the market's concentration (and the lower its competition). As we will explain later, this measure will be useful in order to detect some potential rug pulls in Uniswap. Below, we provide a mathematical definition of the HHI.
\begin{definition} Let $T$ be a token, $\mathcal A$ be the set of addresses and $B_t:\mathcal A\rightarrow \mathbb R_{\geq0}$ be the balance mapping in some time frame $t$. 
The Herfindahl-Hirschman Index of the token $T$ in time $t$ is defined as 
\begin{linenomath*}
\begin{equation*}
HHI_t := \frac{\sum_{a\in \mathcal A}B_t(a)^2}{(\sum_{a\in \mathcal A}B_t(a))^2}.    
\end{equation*}
\end{linenomath*}
The $HHI$ curve is defined as $HHI:[t_{init},t_{end}]\rightarrow [0,1]$.
\end{definition}
 In the cryptocurrency ecosystem, decentralization and the proper distribution of resources are  important features. A smoother distribution of power implies less risk of breakdowns or malfunctioning produced by some participants misbehaving. In exchanges, a similar pattern occurs. In general, the more centralized the capital or funds, the higher the risk of market manipulations or liquidity removal, implying loss of funds by retail investors.\newline
 From a  game theory perspective, the more uniformly distributed the tokens and the liquidity, the less likely it is that agents can manipulate the market or remove funds in a short time period. For this reason, the lower the HHI, the better for the investors. Clearly, one of the problems of the HHI is that it is easy to manipulate and is sensitive to Sibyl attacks. Since any adversary can create an arbitrary number of addresses and transfer an arbitrary amount of tokens among the
addresses on his control. However, these manipulations incur some costs for the malicious agent, reducing the net profits of the attack.

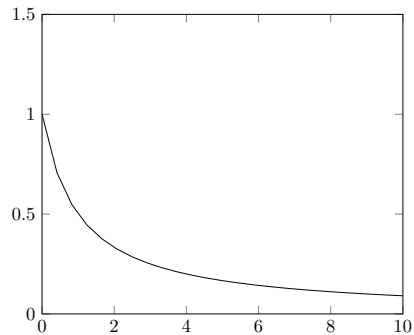
\begin{figure}[!h]
    \centering
    \begin{tikzpicture}[scale =0.7]
    \begin{axis}[
    xmin = 0, xmax = 10,
    ymin = 0, ymax = 1.5]
    \addplot[
        domain = 0:10,
    ] {1/(x+1)};
    \end{axis}
    \end{tikzpicture}
    \caption{Ideal Herfindahl-Hirschman curve}
    \label{Ideal_curve}
\end{figure}
To mitigate some of the problems mentioned, we propose using deeper network analysis tools.
Recent studies used token transactions as a tool to forecast prices \cite{van2018alternative,}, detect price anomalies \cite{li2020dissecting} and detect possible malicious activities \cite{ijcai2020-621,oforiboateng2021topological,patel2020graph}. Each period will be considered as a set of blocks with a separation of approximately one day between them.
\subsubsection{Transaction graph analysis}
The set of transactions and transfers provides insightful information, however this set can be better studied giving it the natural graph structure, interpreting each address as a node and each transaction  as a weighted edge.
\begin{definition}
(Transaction graph) Let $G = (V, E,w)$ be a directed and weighted graph where $V$ is the set of unique addresses, $E \subseteq V \times V$ the transfers from one address to another, and $w(\texttt{tx})$ the amount transacted by the transfer $\texttt{tx}$.
Then,  $G = (G_{1}, G_{2}, \dots, G_{n})$, $i= 1, \dots, n+1$ is the time series for which $G_{i}$ represents the transaction graph generated during period $i, i+1$.
\end{definition}

This time series captures the relationship between the users at each period. This allows us to  study the circulation of the token between the different addresses.

Now, for each $ G_ {i}=(V_i, E_i),$ we define the number of transactions $N_{tx}=\#E_i$, the number of unique addresses $N_{addr}= \#V_i$ and the volume transacted as $\sum_{e\in E}w(e)$. Finally, the average clustering coefficient is defined as $ACC_{i} = \frac{1}{n}\sum_{u\in V_{i}} c_{u}$ where $c_{u}$ is the geometric average of the subgraph edge weights \cite{2005}, computed as follows
\begin{linenomath*}
\begin{equation*}
    c_{u} = \frac{1}{deg(u)(deg(u)-1)} \sum_{vw\in E} (\hat{w}_{uv}\hat{w}_{vw}\hat{w}_{wu})^{1/3}
\end{equation*}
\end{linenomath*}
and $\hat{w}_{uv}$ are normalized by the maximum weight in the network, that is  \newline $\hat{w}_{uv} = w_{uv}/\max_{uv\in E}(w(uv))$.

The average clustering coefficient is a measure of network segregation that captures the connections of individual nodes and their neighbors. In our scenario, this calculation allows us to quantify the interaction of each of the addresses with their neighboring addresses in a given period of time.
\begin{figure}[!h]
\centering
    \begin{tikzpicture}
    \def \n {20}
    \def \N {8}
    \def \radius {3cm}
    \def \rd {1mm}
    \def \rer {4mm}
    
    \def \margin {8} 
    
    \node[draw, circle] at (360:0mm) (ustar) {$\texttt{Pool}$};
    \foreach \i [count=\ni from 0] in {0,1,2,3,4}{
      \node[draw, circle] at ({108-\ni*23}:\radius) (u\ni) {$\texttt{add}_{\i}$};
      \node at ({115-\ni*23}:\radius/2) {$w_{\i}$};
      \draw (ustar)--(u\ni);
    }
    
    \foreach \i in {1,3,...,11}{
      \node[circle] at ({-\i*23}:\radius) (aux) {\phantom{$u_{5}$}};
      \draw[dotted, shorten >=3mm, shorten <=3mm] (ustar)--(aux);
    }
    
    \draw[dotted,red] (18:\radius/2) arc[start angle=18, end angle=-226, radius=\radius/2];
    \end{tikzpicture}
    \caption{Transaction graph of centralized token, with average clustering coefficient 0.}
    \label{fig:star_graph}
\end{figure}
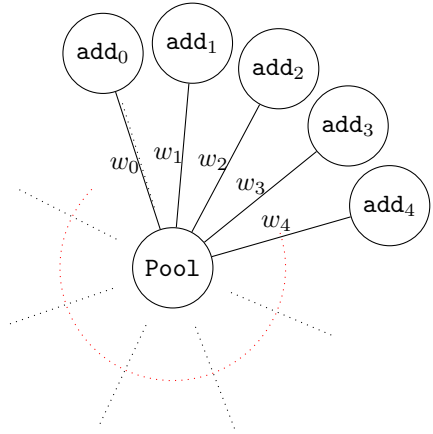

In general, the lower the use case of the token, the lower the average cluster coefficient. 
The main reason behind this heuristic is that low diameter and use case usually implies that the transaction graph is close to a star graph (figure \ref{fig:star_graph}) with \texttt{Pool} being the center node. Therefore, the average cluster coefficient is close to zero.
However, users use non-malicious tokens in a large range of protocols, causing the nodes of these protocols to have a non-trivial cluster coefficient.
Moreover, the daily average cluster coefficient is more expensive to bias due to the constant need to make transaction between \textit{sybil} nodes and the impossibility of using \texttt{batchTransfer} operations, a type of operation that allows assets to be transfered to different addresses in one transaction.


\newpage
\section{Malicious Uniswap Maneuvers}\label{section:token_taxonomy}
As we have mentioned before, a rug pull is one of the most popular ways of scamming in decentralized exchanges combined with a phishing attack. Different techniques are used to trick new investors into buying  malicious tokens. 
To understand the malicious tokens traded in UniswapV2, we introduce a comprehensive classification by manually classifying both malicious and non-malicious tokens. While this classification will provide a clear overview of the tokens in Uniswap, it depends on non-observable variables, such as intentionality and profits. Therefore, we will use a weaker classification for our machine learning model. In this section, we will propose an ideal classification that will provide insights into different rug pulls. Figure \ref{fig:taxonomy} provides an overview of the classification.
\tikzset{
  basic/.style  = {draw, text width=2cm, drop shadow, rectangle},
  root/.style   = {basic, rounded corners=2pt, thin, align=center, fill=white},
  level-2/.style = {basic, rounded corners=6pt, thin,align=center, fill=white, text width=3cm},
  level-3/.style = {basic, thin, align=center, fill=white, text width=1.8cm}
}

\begin{figure}[!h]
\ifadjust
\begin{adjustwidth}{-2cm}{0cm}
\fi
\begin{tikzpicture}[
  level 1/.style={sibling distance=12em, level distance=5em},
  edge from parent/.style={->,solid,black,thick,sloped,draw},
  edge from parent path={(\tikzparentnode.south) -- (\tikzchildnode.north)},
  >=latex, node distance=1.2cm, edge from parent fork down]

\node[root] {\textbf{Malicious Uniswap Maneuvers}}
  child {node[level-2] (c1) {\textbf{Rug Pulls}}}
  child {node[level-2] (c2) {\textbf{Pump-and-dump Schemes}}}
  child {node[level-2] (c3) {\textbf{Money Laundering}}}
  child {node[level-2] (c4) {\textbf{Others}}};
\begin{scope}[every node/.style={level-3}]
\node [below of = c1, xshift=10pt] (c11) {Simple rug pull};
\node [below of = c11] (c12) {Sell rug pulls};
\node [below of = c12] (c13) {SC Trap doors};
\end{scope}
\foreach \value in {1,2,3}
  \draw[->] (c1.195) |- (c1\value.west);

\end{tikzpicture}
\ifadjust
\end{adjustwidth}
\fi
    \caption{Malicious Uniswap Maneuvers classification.}
    \label{fig:taxonomy}

\end{figure}
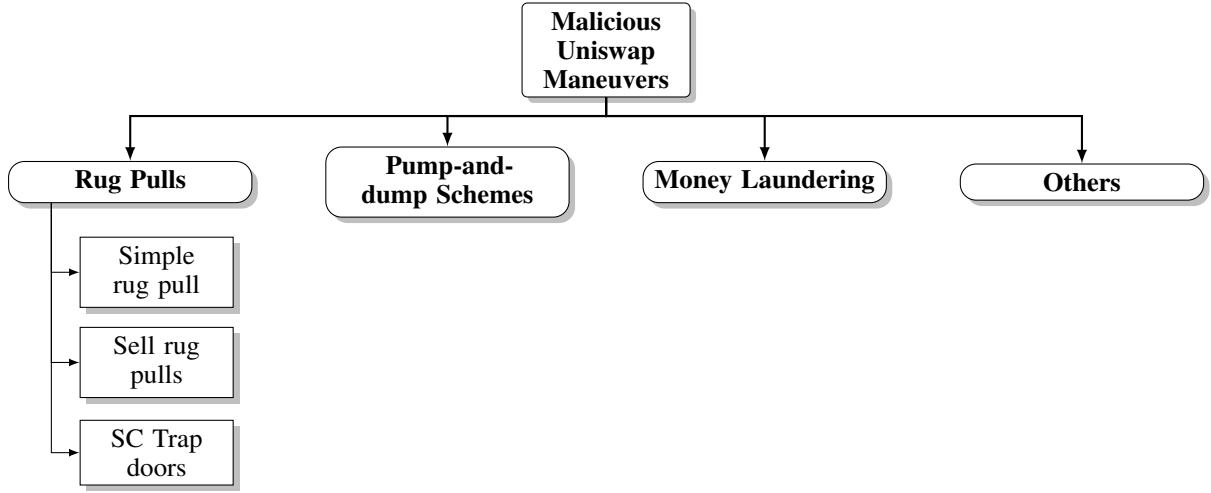

The terms scam and malicious token are not being used identically by all researchers. For example, papers such as \cite{demystifying} use  the terms scam and under-performing token indistinguishably, leading to inaccurate results and classifications. In the current paper, we define a \emph{malicious token} as one released by a developer or a group of developers with no intrinsic value or use case. This definition is similar to that used in paper \cite{liebau2019crypto}. Clearly, this definition is ambiguous unless value and use case are properly defined. In general, this is a complex issue to answer. For example, it is not clear that cryptocurrencies such as Doge\footnote{https://dogecoin.com/} or Shiba \footnote{https://www.shibatoken.com/} have any use case or intrinsic value, but they are among the most popular meme-coins. 
In our framework, we say that a token has no intrinsic value or use case if the developer knows that the trading price with respect to USD  will eventually be zero. In other words, a tradable malicious token in Uniswap induces a zero-sum game between the users and the developers, i.e. the incentives for the investors are not aligned with those of the token creators. Therefore, the main difference between malicious and non-malicious tokens is the developer's intentionality towards the token. One of the main problems of these definitions is that it is unfeasible to distinguish between scam tokens and under-performing or abandoned projects without accurate off-chain data. However, in general, the two terms do coincide.  For these reasons, in the following section, rather than trying to identify whether a token is malicious or not, we will give a methodology for classifying and predict under-performing and inactive tokens. 

\textit{Simple rug pulls} are the most common and easy to identify rug pulls. Essentially, simple rug pulls consist of  three steps. The developer creates an ERC-20 token $\mathcal C$ and interacts with UniswapV2 Factory to create a new trading pair with wETH  or any other relevant token, fixing the reserves to $(x,y)$. Then, the investors execute the swap transaction on the trading pair, exchanging ETH for $\mathcal C$, and the reserves update to $(x+\Delta_x,y-\Delta_y)$. 
Afterwards, the developer activates the function \texttt{removeLiquidity} obtaining $x+\Delta_x$ Ether and $y-\Delta_y$ of $\mathcal C$. 
Since the coin $\mathcal C$ has no intrinsic value, the net profit of the attacker is $x+\Delta_x-\text{fees}$.\newline
\textit{Sell rug pulls} are also prevalent. However, they are not easy to identify and compute the total gains of the attack. A simplification of the attack would be as follows. The developer creates an ERC-20 token $\mathcal C$, with total supply $S$, and   a new trading pair with a relevant token $\mathcal E$. The developer adds a fraction $f<S$ of the total supply of the $\mathcal C$ to the pool, having complete control of the remaining coins $S-f$. Then, they wait for a sufficient number of investors to execute the swap transaction on the trading pair.  Afterwards, the developer swaps $f$ coins $\mathcal C$ for $\mathcal E$. While this kind of rug pull is theoretically less profitable for the attacker, if combined with more features it can be even more profitable than the first one. For example, in order to build confidence in investors, the attackers lock the liquidity in a smart contract or burn it. This makes investors think that a simple rug pull cannot happen, and that therefore it is a potentially profitable investment. In other words, the fact that the liquidity is locked makes the market volume increase. Moreover, if the token is mintable, the attacker can recover all the funds, minting as many coins as needed to recover nearly all the tokens $\mathcal E$ in the trading pool.
\newline
\textit{Smart Contract Trap door rug pull} are the most difficult to identify and prevent. There are several reasons for this. The first is that the EVM is Turing complete, and therefore it is highly complex to identify all  vector attacks; the second is that smart contracts do not exist in isolation, i.e., smart contracts can work on top of other smart contracts. This means that the economic security of some smart contracts, depends on other smart contracts \cite{babel2021clockwork}. In the following section, we will explain some of the most popular examples of these trap-doors.

\begin{itemize}
    \item \texttt{Mintable} is a property shared by many tokens, including non-malicious ones such as USDT\footnote{https://tether.to/}. In general, we say that a token is mintable if it has a function that allows it to increase the supply of the token with some pre-defined conditions. Usually, mintable tokens give rights to mint new  tokens to the developers to a fixed set of addresses. While this functionality can be a useful feature in some contexts, it can also be used by malicious users to subtract all the liquidity of the pull by minting as many tokens as needed.
    \item \texttt{TransferFrom/Approve} bad design is a property shared by some malicious tokens. \newline \texttt{TransferFrom} is the function that allows a smart contract to transfer assets on behalf of an externally owned account. In the context of  Uniswap, a proper design of this function allows tokens to be sold. 
    In other words, arbitrarily changing the \texttt{transferFrom} function makes the Uniswap Pool behave as a honeypot \cite{torres2019art}. 
    There are different ways of making this smart contract, however the most popular example of this kind of scam is the use of tokens that contain the line code \texttt{require(from == owner $\|\|$ to == owner $\|\|$ from == UNI)} in the \texttt{transferFrom} function.
    \item Composability vulnerabilities are the least common ones and the hardest to find. In general, this type of rug pull is not made by the developers but by malicious agents external to the project, who take advantage of the bad design of the smart contract token interacting with Uniswap or other DEXs. Those that we have been able to identify are the tokens with a price oracle vulnerability. Usually, we observed that these tokens have Uniswap integrated in the source code of the smart contract in order to reward holders. However, these rewards depend on the price. The higher the price, the higher the reward. The fundamental problem of this mechanism is that the price is defined through an oracle that, as shown in \cite{eskandari2021sok}, is easy to manipulate with enough funds or using a flash loan.
\end{itemize}
In general, these rug pulls are not exclusive and in reality, different techniques are applied in order to execute a rug pull. Moreover, these techniques are usually combined with phishing attacks and pump-and-dump schemes.\newline
\textit{Money laundering:} In traditional finance, money laundering is the processing of money obtained from illicit activities, to make it appear that it originated from a legitimate source. 
In the Ethereum environment or cryptospace in general, we will define money laundering as the process of sending some coins obtained from heists, honeypots or hacks from an address $\texttt{addr}_{1}$  to another address $\texttt{addr}_{2}$ privately. That is, observers cannot link these two addresses through the transaction graph. In Ethereum, the most common technique to obtain privacy is through Mixers \cite{seres2019mixeth}. 
The most popular mixer is TornadoCash \footnote{https://tornado.cash/}, which implements a smart contract that accepts transactions in Ether so that the amount can later be withdrawn with no reference to the original transaction by means of using zero-knowledge proofs. 
Analyzing rug pulls in the Uniswap protocol, we have found some users that use the Uniswap protocol to send Ether to other address without being noticed by most common address clustering algorithms. 
The operation consists of three main steps. First, the address $\texttt{addr}_{2}$ creates an unsellable token $\mathcal C$ and lists it in Uniswap with an arbitrary amount of liquidity. 
Then, the address $\texttt{addr}_{1}$ changes the Ether for $\mathcal C$ in the Uniswap pool. Finally, the address $\texttt{addr}_{2}$ removes all the liquidity from the pool.\newline
\begin{figure}[!h]
    \centering
    \includegraphics[scale = 0.75]{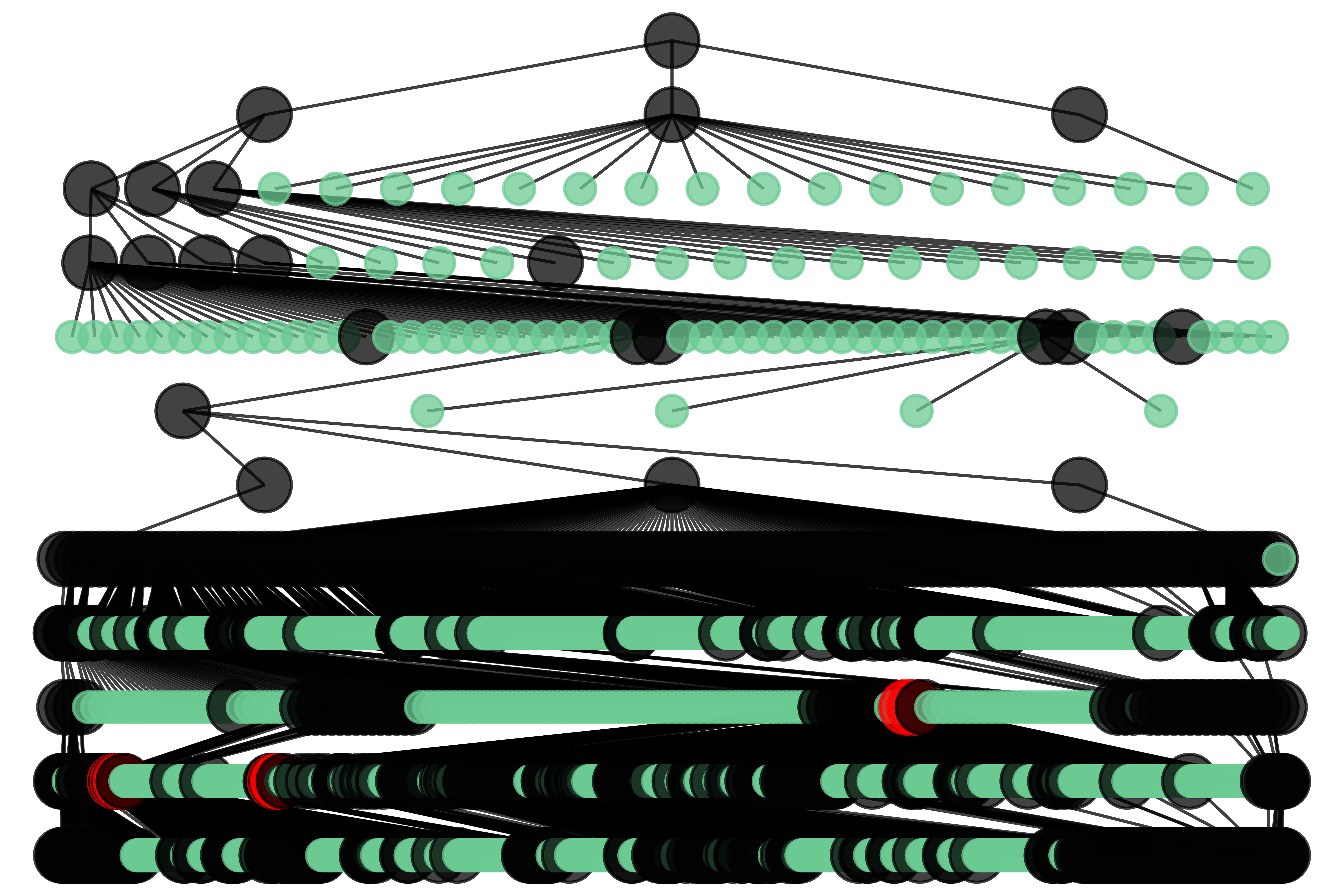}
    \caption{
    Transaction graph with depth=10, of scammer \texttt{addr} = 0x775744...\protect\footnote{https://etherscan.io/address/0x775744a529f73a754164e4fE740e44C7c5aa5942}, where black nodes are addresses directly connected, green nodes are transactions of type rug pull/money laundering and red nodes are centralized exchanges, such as Binance, Coinbase and Huobi.
    }
    \label{money_laundering}
\end{figure}
A good example of this is given by
the user that controls the address 0x775744...\protect\footnote{https://etherscan.io/address/0x775744a529f73a754164e4fE740e44C7c5aa5942}, in charge of creating more than $500$ tokens in order to money launder, pump prices and obtain profits via executing trap door rug pulls.\newline
\textit{Pump-and-dump schemes}: In traditional finance, a pump-and-dump scheme is a malicious maneuver that manipulates the market price of a stock, in which the executors first purchase a financial asset at a certain price. They then persuade other speculative non-informed investors to purchase, within a short period of time thereby causing the price rise artificially (pump), and executors sell their assets at profit. This typically leads to a rapid price drop (dump), leaving the victims with a loss. 
In traditional markets, pump-and-dump schemes have generally been illegal around the globe. However, in cryptocurrencies, the lack of regulations and the nature of cryptospace allow these maneuvers to happen easily and avoid sanctions. While these maneuver have some intersection points with the rug pull maneuver, the fundamental difference is that in pump-and-dump schemes the targeted asset is not necessarily malicious, while in rug pulls it is.
\newline
\textit{Others:} While malicious maneuvers are usually a combination of the ones mentioned before, all of them have a characteristic in common, the victims are uninformed users. However, there are other maneuvers in EVM-compatible blockchains that try to attack weaknesses of maximal extractable value (MEV) bots. A good example of this is Salmonella\footnote{https://github.com/Defi-Cartel/salmonella}, a bot that tries to trick sandwich traders \cite{highfrequency}. Salmonella creates a token with an \texttt{approve/transferFrom} bad design. Afterwards, it creates a swap transaction, that tricks MEV bots to sandwich it with a buy and sell operation. At the moment the transactions are executed, by design of the Salmonella smart contract, the buy gets accepted, but the sell transactions is reverted, leaving a lot of cash in the pool for the Salmonella developer.

 \section{Data Collection}\label{section:data_section}
 To download all the data needed to do the labelling and the analysis, we used an Infura archive node  \footnote{https://infura.io/}  and the Etherscan API \footnote{https://etherscan.io/apis}. 
 To obtain the state of the Uniswap exchange and the tokens, we  used the events produced by their respective smart contracts.
 Any node connected to Ethereum JSON-RPC API can observe these events and act accordingly. 
 Events can also be indexed, so that the event history is searchable later.
 \begin{enumerate}
     \item \textit{Tokens listed}: We obtained the history of all tokens listed in the Uniswap V2 from its creation to 03/09/2021 asking for all events of the \texttt{PairCreated} type in the UniswapV2Factory contract.
     \item  \textit{Smart contract and features}: After obtaining all listed tokens in Uniswap, with the help of Etherscan, we downloaded the transactions in which they were created, their smart contract, their decimals, and their symbol. In order to speed up these calls we used the multicall contract \footnote{https://etherscan.io/address/0xb1f8e55c7f64d203c1400b9d8555d050f94adf39} to batch these calls to the blockchain in a single call. Afterwards, we used Slither\cite{feist2019slither} to obtain different features of the smart contract, such as \texttt{pausable} and \texttt{mintable}.
     \item \textit{Events}: From all the pools of Uniswap obtained in \textit{Tokens listed}, we collected all events of type \texttt{Sync, Mint, Burn} and \texttt{Transfer} for each of the PairPools obtained. Finally, we downloaded all \texttt{Transfer} events from each of the tokens.
 \end{enumerate}
 There are several attributes that we could not download via API such as transaction creation of a contract and full market cap of a token, however, they are available on certain block explorers such as Etherscan. In these cases, we have used scrapping techniques to obtain that information. 
 For example, some of the data that we have not been able to find via the Etherscan API and Infura is: the hash of the transaction in which tokens were created and tokens that have had some type of external audit.


\section{Token labelling}\label{section:Class}
In this section, we provide the set of tools and the methodology we used in order to label the tokens listed in the Uniswap protocol as malicious and non-malicious, and provide the results obtained using this methodology. First, we define the maximum drop and the recovery of token prices and liquidity time series. Then, we explain the distribution of tokens that eventually became inactive or that became a rug pull. Finally, we explain which methodology we used to accurate label tokens as non-malicious.
\subsection{Ground truth labelling}

One of the final goals of this study is to create an ML algorithm capable of detecting malicious tokens. To do this, we have built a list of tokens tagged as malicious and non-malicious. In this case, the label of the malicious tokens has been deduced from a series of calculations defined below:
\begin{definition}
Let $X = \{X_{t}\mid t\in\{0,...,S\}\}$, be the time series representing the price or liquidity in all the token activity up to the last sync event $S$. The maximum drop is defined as
\begin{equation*}
MD = \left|\frac{X_{l}-X_{h}}{X_{h}}\right|,    
\end{equation*}
where $X_h = \text{max}\{X_\tau\mid\tau\in\{0,...,S\}\}$, $h = \text{argmax}\{X_\tau\mid \tau\in\{0,...,S\}\}$ and $X_l = \text{min}\{X_\tau\mid\tau\in\{h,...,S\}\} $.
\end{definition}

The maximum drop is usually known in the literature as maximum drawdown and is often used as a risk measure of portfolios. Informally, the maximum drop is the largest drop from a peak to a trough. In our context, the maximum drop measures fall in the price or liquidity of the Uniswap listed pools. In section \ref{section:token_taxonomy}, we have seen that the last step in a simple rug pull is the removal of all liquidity from the pool. Therefore, by definition, in a simple rug pull, the MD of the liquidity or time series of a rug pull tends to be approximately $1$. However, the opposite implication is not true in general. While the price maximum drawdown being close to $1$ implies that the token is malicious, the MD of liquidity being $1$ does not necessarily imply some malicious behavior. 
For example, the developer could have moved the funds to another pool or another DEX project. Moreover, it could be possible that the market maker just wants to retire their funds and does not have any more interest in providing liquidity. In general, if the token has a use case and a market value, other agents will have incentives to take over as market makers. For this reason, we introduce the recovery.

\begin{definition}\label{def:recovery}
Let $h, l \in [0, S]$ be the elements defined previously. Then, the recovery from $X_{\tau}$ to $X_{S}$ is computed as 
\begin{equation*}
    RC = \frac{X_{S} - X_{l}}{X_{h} - X_{l}}.
\end{equation*}
\end{definition}
\begin{figure}[H]
    \centering
    \includegraphics[scale=0.6]{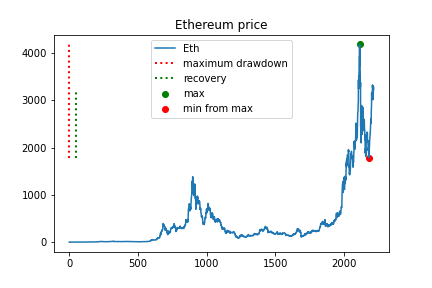}
    \caption{Maximum drawdown and recovery of ETH price,}
    \label{fig:MD}
\end{figure}
Informally, the recovery is the largest pump from the bottom. This measure makes it possible to check if the liquidity position and the price of a token have recovered after the drop. Next we will show how the data splits, taking into account the maximum drawdown and the recovery.
\begin{figure}[H]
        \centering
        \begin{subfigure}[b]{0.355\textwidth}
            \centering
            \includegraphics[width=\textwidth]{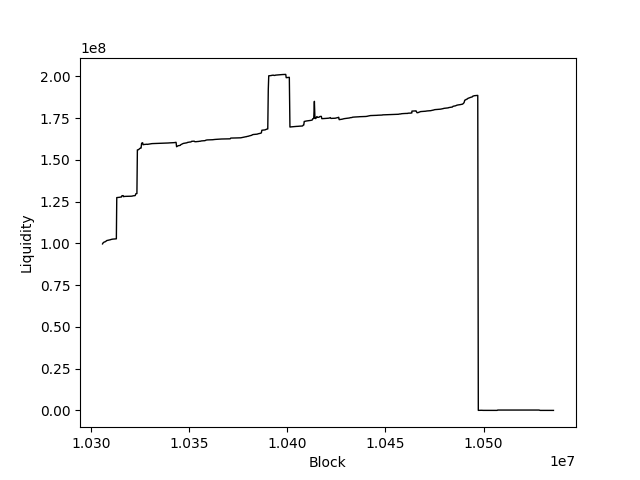}
            \caption[Network2]%
            {{\small Liquidity Fast Rug Pull}}    
            \label{fig:mean and std of net14}
        \end{subfigure}
        \begin{subfigure}[b]{0.355\textwidth}
            \centering 
            \includegraphics[width=\textwidth]{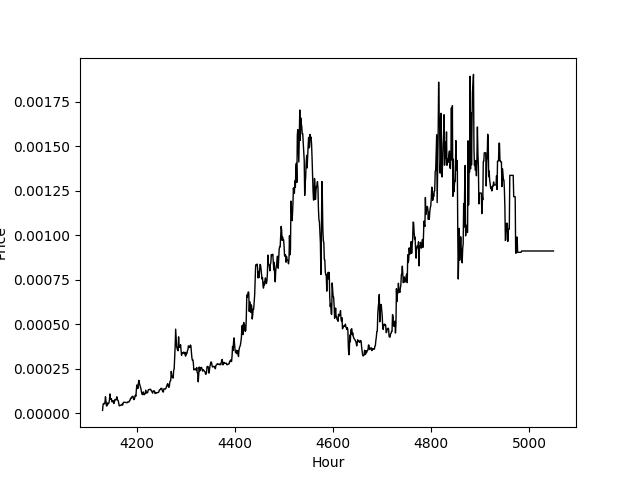}
            \caption[]%
            {{\small Price Fast Rug Pull}}    
            \label{fig:mean and std of net24}
        \end{subfigure}
        \vskip\baselineskip
        \begin{subfigure}[b]{0.355\textwidth}
            \centering 
            \includegraphics[width=\textwidth]{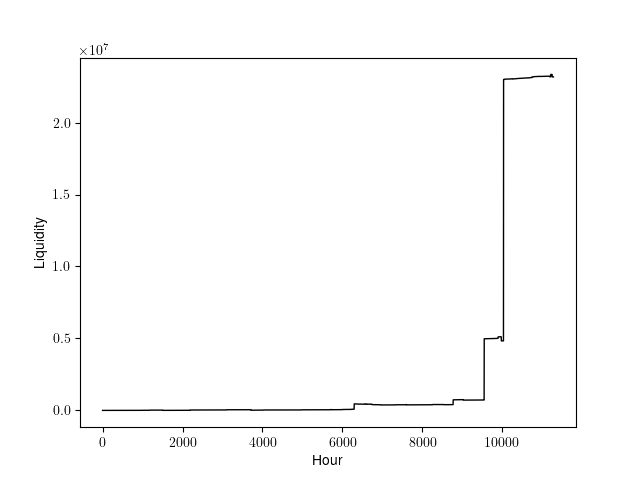}
            \caption[]%
            {{\small Liquidity Rug Pull without Burn Events}}    
            \label{fig:mean and std of net34}
        \end{subfigure}
        \begin{subfigure}[b]{0.355\textwidth}  
            \centering 
            \includegraphics[width=\textwidth]{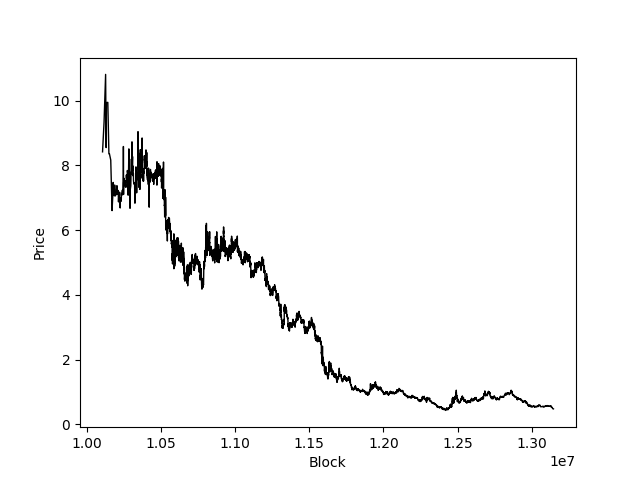}
            \caption[]%
            {{\small Price Rug Pull without Burn Events}}    
            \label{fig:mean and std of net44}
        \end{subfigure}
        \caption[]
        {This Figure shows the price and liquidity time series of two different types of rug pulls. The two first pictures are associated with a simple rug pull with token 0x896a07e3788983ec52eaf0F9C6F6E031464Ee2CC, while the second pair of pictures are associated with a sell rug pull with token  0x0A7e4D70e10b63FeF9F8dD19FbA3818d15154d2Fa.} 
        \label{fig:mean and std of nets}
    \end{figure}
The figure \ref{fig:mean and std of net14} shows two examples of different rug pulls with no recovery and maximum drop of one in liquidity and price respectively. Moreover, one can induce that the rug pull $a)$ is a simple rug pull while the rug pull $b)$ is a sell rug pull.
\subsubsection{Malicious Tokens Labelling}\label{subsection:malicious_tokens}
Various features were computed, taking into account two properties: fluctuations in price or liquidity, and activity. As explained above, the maximum drop computes the greatest drop, either in liquidity or price, during the activity of the tokens. 

Most malicious tokens, at some point, lose all their liquidity or their price drops to zero. However, this does not necessarily indicate malicious behavior, as it may be due to a simple fluctuation. Thus, we also compute the recovery. If a token loses all its liquidity or its price drops to zero and these levels are never recovered, then the probability that the falls are due to malicious intent increases. 

In addition, to ensure that these fluctuations are not due to simple market movement, we compute the time elapsed since the last movement of the pool or token transfer to 13/09/2021. If more than one month has passed between the last movement or transaction of the token so far, we consider that the token is inactive. Finally, we obtain a list of inactive tokens, which have drops in price or liquidity of almost one hundred percent and which do not recover.
Our initial list contained 46,499 tokens. We discarded those that did not have decimals defined in their contract (169).
We then selected those that had a pool connected to wETH (44,685) and downloaded their \texttt{Sync, Mint, Burn} and \texttt{Transfer} events.
The final list contains 37,891 tokens that have at least one pool connected with wETH and more than 5 Sync events in all their activity. This last property is necessary to be able to compute the heuristics used to label the tokens.

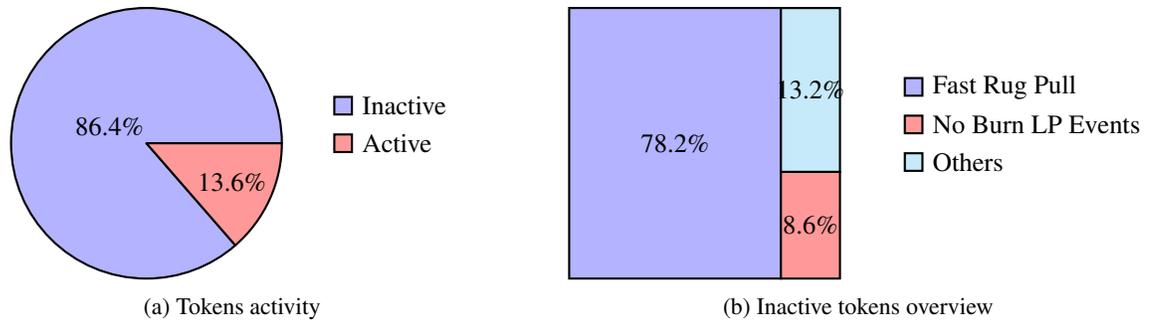
\begin{figure}[H]
\ifadjust
\begin{adjustwidth}{-0.5cm}{0cm}
\fi
\hspace{0.32cm}
    \begin{subfigure}{.5\textwidth}
        \centering
        \begin{tikzpicture}[scale=0.60]
            \pie[color = {blue!30,red!40},text = legend]{86.4/Inactive, 13.6/Active}
        \end{tikzpicture}
        \caption{Tokens activity}
    \end{subfigure}
    \begin{subfigure}{.5\textwidth}
        \centering
        \begin{tikzpicture}[scale=0.60]
        \pie[color = {blue!30,red!40,cyan!20},square,text = legend]{78.2/Fast Rug Pull, 8.6/No Burn LP Events, 13.2/Others}
        \end{tikzpicture}
        \caption{Inactive tokens overview}
    \end{subfigure}
    \caption{Pie charts of features activity (left) and maximum drop (right) for the final list of  37,891 tokens.}
\ifadjust    
\end{adjustwidth}
\fi
\end{figure}

As explained above, we first checked if the tokens were inactive, i.e, if they had not registered \texttt{Transfer} or \texttt{Sync} events for more than 30 days (86.4 \% of the total). We also computed the maximum liquidity drop and saw that the liquidity of 78.2 \% of the inactive pools had been completely withdrawn at some point. Finally, we noticed that only 0.4\% of pools that at some point had lost all their liquidity, recovered in all subsequent activity. This made a total of 24,870 tokens that could be tagged as malicious since they were inactive tokens, that had, at some point, lost all their liquidity and had not recovered it again.
\begin{figure}[H]
    \centering
    \pgfplotstableread[col sep=comma,header=false]{
    $RC = 0$,89.1,60.7
    $0<RC<0.1$,10.5,37.4
    $0.1\leq RC\leq1$,0.4,1.9
    }\data
    
    \pgfplotsset{
    percentage plot/.style={
        point meta=explicit,
    nodes near coords align=vertical,
        yticklabel=\pgfmathprintnumber{\tick}\,$\%$,
        ymin=0,
        ymax=100,
        enlarge y limits={upper,value=0},
    visualization depends on={y \as \originalvalue}
    },
    percentage series/.style={
        table/y expr=\thisrow{#1},table/meta=#1
    }
    }
    
    \begin{tikzpicture}
    \begin{axis}[
    axis on top,
    width=10cm,
    ylabel=Tokens Percentage,
    xlabel=Recovery $RC$ \ref{def:recovery}of Liquidity and Price,
    percentage plot,
    ybar=0pt,
    bar width=0.75cm,
    enlarge x limits=0.25,
    symbolic x coords={$RC = 0$, $0<RC<0.1$,$0.1\leq RC\leq1$}, 
    xtick=data
    ]
    \addplot table [percentage series=1] {\data};
    \addplot table [percentage series=2] {\data};
    \legend{Liquidity,Price}
    \end{axis}
    \end{tikzpicture}
    \caption{Price and Liquidity recover}
    \label{fig:price_liquidity}
\end{figure}
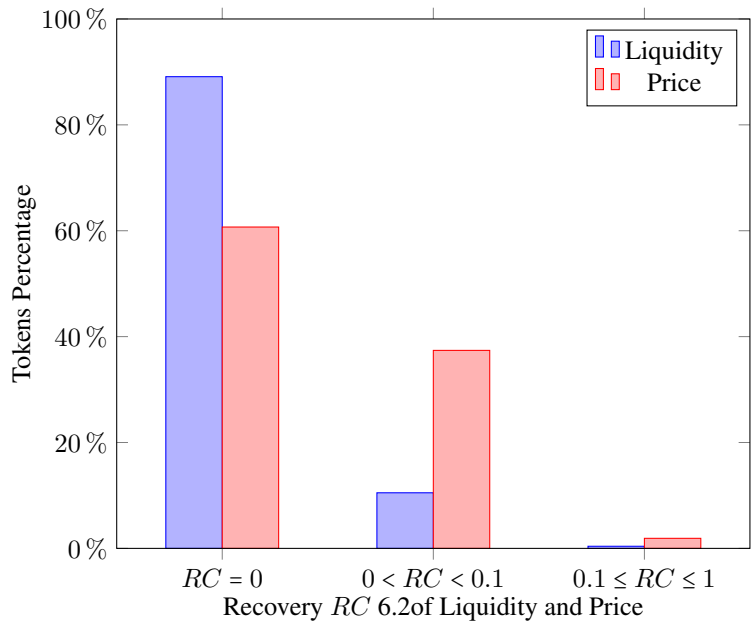
On the other hand, as shown in Figure \ref{fig:price_liquidity} 8.6\% of the inactive tokens, did not have any Burn LP events in all their activity period. However, 79.2\% of this 8.6\% had seen a price drop of more than 90\% at some point, and only 1.9\% recovered their value after the drop. This adds 2,087 tokens that can be identified as malicious since they are inactive, with price MD of at least 90\% and no revovey.

\subsubsection{Non-Malicious Tokens Labelling}
Unlike malicious tokens, non-malicious tokens cannot be chosen from a liquidity, price, and activity analysis. Given a token, it may be considered malicious if there has been at least one rug pull at some point in its activity. However, a token that has not had any rug pull cannot be considered non-malicious, since it could experience a rug pull later on. Therefore, we take advantage of audits carried out by external companies (Certik, Quantstamp, Hacken...). It is important to highlight that non-malicious tokens can have drops in price and liquidity, too. Nevertheless, none of them simultaneously fulfil all the three properties that define malicious tokens, namely: inactivity, a sharp drop in price or a sharp drop in liquidity, and no recovery. Thus, a list of 674 tokens labelled as non-malicious have been mined from different sources \footnote{https://coinmarketcap.com/view/defi/\newline https://www.coingecko.com/en/categories/decentralized-finance-defi \newline
https://etherscan.io/tokens}. We have also discarded those that are so large that it becomes computational expensive to compute its features, for example USDT or USDC. The final list contains 631 tokens labelled as non-malicious.

\section{Scam Detection}\label{section:Scam}
We start from a list of tokens labelled as malicious or non-malicious, according to their features, therefore it can be considered a binary classification problem. In this classification, we distinguish between the two types of tokens in the moments prior to the malicious activity. This means that models are capable of detecting malicious tokens in the activity prior to the rug pull.
In this section, we present two methods: one considering all the activity of a token, and the other considering just the first 24 hours. Also, we detail the different classifiers used in both methods and their hyperparameter optimization. Finally, we present the results of each method.

\subsection{Activity based Method}

Our goal is to detect malicious tokens at an early stage, i.e., before users lose their capital. So far, we have characterized two main types of rug pull: the ones that lose all liquidity at some point and the ones where the price drops to almost zero. In this way, for each token labelled as malicious, we have randomly chosen several evaluation points prior to the maximum drop. Non-malicious tokens have been evaluated throughout their activity. Then, for each evaluation point, we have calculated the token features up to that block and used them to train two ML algorithms in order to find those patterns related to malicious activity.

\begin{figure}[h]
\ifadjust
\begin{adjustwidth}{0cm}{0cm}
\fi
  \subfloat[Liquidity]{\includegraphics[width=0.45\textwidth]{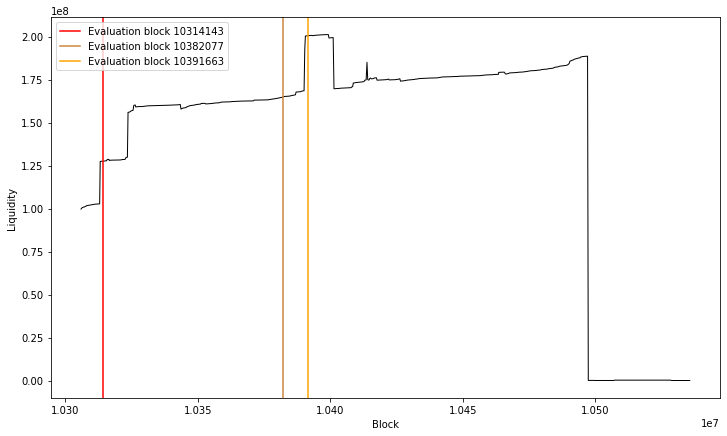}\label{fig:liquidity}}
    \qquad
  \subfloat[Price]{\includegraphics[width=0.45\textwidth]{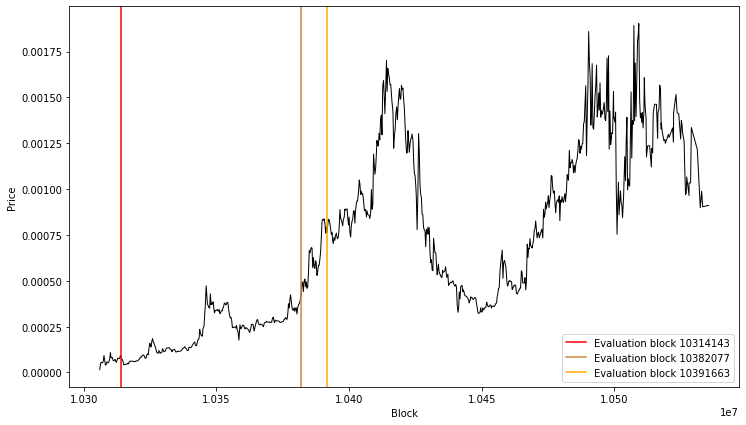}\label{fig:price}}
  \caption{Evaluation points in liquidity (left) and price (right) chosen for token 0x896a07e3788983ec52eaf0F9C6F6E031464Ee2CC labeled as malicious. \newline}
  \label{fig:liq_price}
\ifadjust
\end{adjustwidth}
\fi
\end{figure}

In this method, we choose $n$ evaluation blocks at random. For example, Figure \ref{fig:liq_price} shows the price and liquidity of one token labelled as malicious. In this token, liquidity suddenly drops to zero and does not recover again. Therefore, we consider it to be a fast rug pull. The three vertical lines leading up to the crash represent the three evaluation points for that particular token. This means that we have calculated the variables of that token up to those blocks. In this way, we have proceeded with each of the labelled tokens. As explained in section \ref{subsection:Results}, the final metrics have been computed taking five evaluation points on non-malicious tokens and one on malicious tokens. 
All evaluations are prior to the malicious act; this implies that this method can later be used as a tool to detect malicious tokens at any time. However, there are subtleties that can skew the ML algorithms used. For example, tokens labelled as non-malicious tend to have a much larger capitalization compared to malicious tokens, therefore the algorithm could end up differentiating between "small" tokens and "large" tokens instead of malicious and non-malicious. Although this differentiation is not a bad approach to this task, we think that there may be other situations that require different approaches. In the next method, we evaluate all tokens at the same temporal evaluation points in order to identify these possible biases.

 \subsection{24 Early Method}
Rug pulls are profitable if their malicious act is done before they are discovered. Therefore, most rug pulls (93\%) occur in the first 24 hours after the pool is created. This encourages us to build a tool to detect malicious tokens at startup. For each labelled token, we have computed its features in each of the 24 hours after its pool creation. Then, we create a different dataset for each hour in which the tokens are evaluated. 

Note that, in this case, we are training the models for each hour, therefore, we only have one evaluation point for each dataset. This also implies that we will have a smaller dataset compared to the other method. In Section \ref{section:ResultsSecondMethod}, we present the different metrics obtained for each of the hours. Also, we can measure the evolution of the predictive power of the algorithm in the first hours of the token's life. The fact that this method requires a certain history to be able to give a prediction implies that the more history you obtain, the better the prediction. This intuition is confirmed in \ref{subsection:Results}.


 
 \subsection{Machine Learning and Hyperparameter Optimization}
 
\textit{Gradient Boosting Decision Tree} (GBDT)\cite{gbdt} models offer high performance in classification tasks with tabular data since they allow the definition of different cost functions, do not require preprocessing for categorical features, and can handle missing data. Thus, it seems clear to apply a GBDT model to our problem. In particular, we have used the \textit{XGBoost} \cite{xgboost} algorithm.

On the other hand, algorithms with Transformer architecture \cite{Transformer} are obtaining high results in fields such as natural language processing \cite{BERT, ROBERTA}, computer vision \cite{TransformerVision}, etc. In this work, we have used a model based on attention mechanisms called FT-Transformer \cite{FT-Transformer} in order to test a tabular algorithm with Transformer architecture in our problem.

Hyperparameters cannot be learned during the training process. Furthermore, they have a significant impact on the performance of the model being trained. Thus, optimizing them is crucial for better efficiency, faster convergence, and overall better results.
In this work, we have used Optuna \cite{optuna}, a software framework designed primarily for hyperparameter optimisation in ML algorithms.

Finally, in order to evaluate the impact of each variable, we have used the SHAP (SHapley Additive exPlanations) Values \cite{SHAP}. SHAP uses game theory to explain the results obtained in ML algorithms. In particular, it uses the classical Shapley values of game theory and their related extensions.

\subsection{Results}\label{subsection:Results}

The final list contains 27,588 labelled tokens,  631 labelled as non-malicious tokens and 26,957 labelled as malicious. Within the malicious, 24,870 are fast rug pulls and 2,087 do not contain LP Burn events.
We see that there are far fewer non-malicious tokens than malicious ones. There are many techniques\footnote{https://imbalanced-learn.org/stable/references/index.html} to deal with this problem, however, none of them have been applied in order to make the results more understandable. Instead, our data augmentation technique consists of choosing more evaluation points for non-malicious tokens than for malicious tokens. Now, given this dataset,  we want to increase the performance in predicting non-malicious tokens since it would be enough to label all of them as malicious to achieve an accuracy of 97,7\%. Therefore, we label the non-malicious tokens as 1 and the malicious tokens as 0. 

We have used the cross-validation method to validate both ML algorithms. Cross validation is a resampling method that uses different parts of the data to test and train a model in different iterations. In particular, we have used the stratified version, in which the partitions are selected so that the mean response value is approximately the same in all partitions. In the case of binary classification, this means that each partition contains roughly the same proportions of the two types of class labels. In the first method, we have five evaluation points on non-malicious tokens and one on malicious tokens. Thus, in each of the iterations, the tokens of the training and validation set are separated in a stratified way with all their corresponding evaluations. This implies that the same token will never have evaluations in the training and validation set at the same time, and all folds will have roughly the same number of malicious tokens. Finally, we used 5-fold cross-validation, therefore all the results will be presented as the mean and standard deviation of all folds.

We have used \texttt{xgboost}\footnote{https://xgboost.readthedocs.io/en/stable/} Python library to apply the XGBoost model to each method. Specifically, in each of the five folds, we have used the training partition to perform a  hyperparameter's optimization (see appendix \ref{section:Hyperparameters}) to later predict the test of the corresponding fold. In the case of FT-Transformer, we have used the default parameters of  \texttt{rtdl}\footnote{https://yandex-research.github.io/rtdl/stable/index.html} Python library, since training this model is too expensive to perform hyperparameter optimization. 

\subsubsection{Activity based Method Results}
Both XGBoost and FT-Transformer get high metrics for accuracy, recall, precision, and F1-Score. However, XGBoost outperforms FT-Transformer in all metrics. As we previously said, unlike XGBoost, FT-Transformer hyperparameters have not been optimized due to the high computational complexity required to train the model.

\begin{table}[H]
\ifadjust
\begin{adjustwidth}{0cm}{0cm}
\fi
\begin{subtable}[c]{0.4\textwidth}
\centering

\begin{tabular}{lrr}
\toprule
{\textbf{XGBoost}} &   Mean &    Std \\
\midrule

Accuracy    &  0.9936 &  0.0029 \\
Recall &  0.9540 &  0.0297 \\
Precision   &  0.9838 &  0.0056 \\
F1-score          &  0.9684 &  0.0151 \\
\bottomrule
\end{tabular}
\subcaption{XGBoost metrics.}
\end{subtable}
\begin{subtable}[c]{0.4\textwidth}
\centering
\begin{tabular}{lrr}
\toprule
{\textbf{FT-Transformer}} &    Mean &     Std \\
\midrule
Accuracy    &  0.9890 &  0.0036 \\
Recall      &  0.9180 &  0.0363 \\
Precision   &  0.9752 &  0.0109 \\
F1-Score    &  0.9454 &  0.0187 \\
\bottomrule
\end{tabular}   
\subcaption{FT-Transformer metrics.}
\end{subtable}
\caption{Accuracy, recall, precision and f1 score obtained in a 5-fold cross-validation for first method.}
\ifadjust
\end{adjustwidth}
\fi
\end{table}

To understand the impact of each feature in both models, we have computed the SHAP values. In this work we only focus on XGBoost, however, the process would be the same for FT-Transformer.
The SHAP values assign the importance of each feature for each prediction. In general, the greater the impact of features on a prediction, the greater the SHAP value in absolute value.

\begin{figure}[H]
\ifadjust
\begin{adjustwidth}{0cm}{0cm}
\fi
  \subfloat{\includegraphics[width=0.45\textwidth]{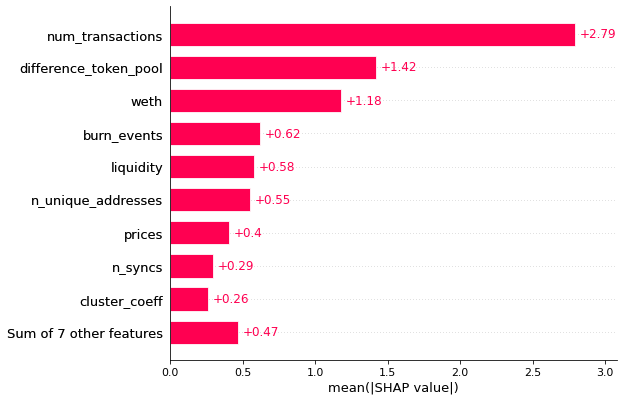}\label{fig:f1}}
    \quad
  \subfloat{\includegraphics[width=0.49\textwidth]{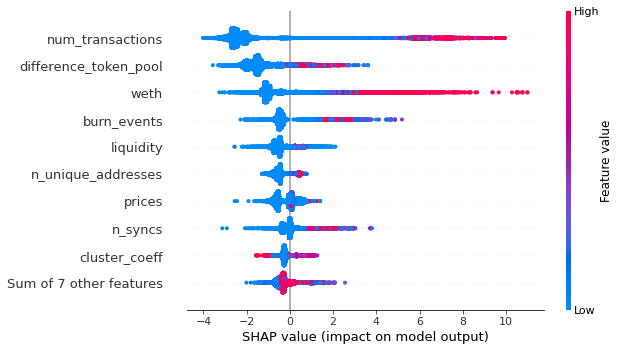}\label{fig:f2}} \caption{Impact of the variables in the XGBoost model applied to the first method.  Mean of  SHAP values on the left and global impact of the features on the right. Images generated from the Python SHAP library.}
  \label{fig:shapvalues}
\ifadjust
\end{adjustwidth}
\fi

\end{figure}

In Figure \ref{fig:shapvalues} we show, on the left side, the feature importance in terms of SHAP value applied in the first method, and, on the right, the impact on the final output. As previously said, most of malicious tokens die in the first 24 hours after the pool is created; by contrast, non-malicious tokens have longer life. This explains why features such as number of transactions or number of unique addresses have so much weight in the model. Another important feature is the difference in blocks between the creation of the token and the pool. We notice that less block difference between token and pool creation implies negative SHAP values, and negative SHAP values should correspond to malicious tokens. This conclusion coincides with \cite{demystifying} since several of the malicious tokens take advantage of social trends by copying the name of official tokens and take money from investors who get confused. This technique implies speed in the creation of the token and the pool, since otherwise the trend may be lost.

\subsubsection{24 Early Method Results}
The results of the second method must be understood from another perspective, since the problem posed is not the same. As we said, the difference with respect to the first method lies in the fact that, we evaluate all the tokens at a certain time after the creation of their respective pools. 

\begin{figure}[H]
\ifadjust
\begin{adjustwidth}{0cm}{0cm}
\fi
  \subfloat[XGBoost]{\includegraphics[width=0.45\textwidth]{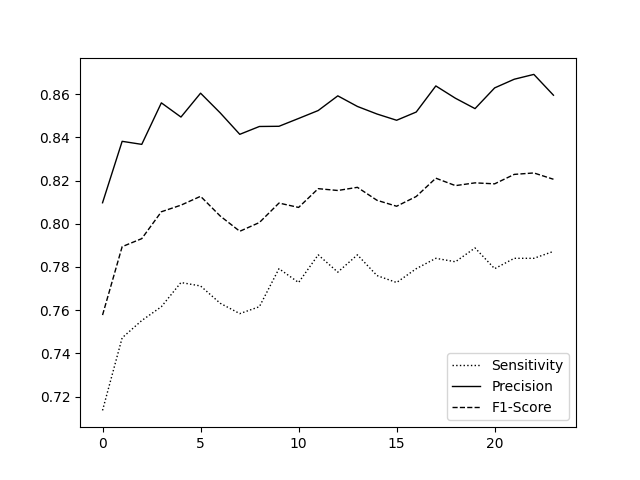}\label{fig:F1score}}
    \quad
  \subfloat[FT-Transformer]{\includegraphics[width=0.45\textwidth]{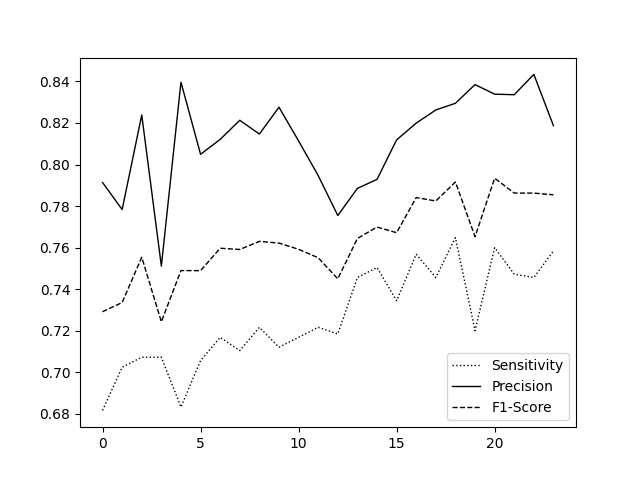}} \caption{Evolution of recall, Precision and F1-score for the XBGoost and FT-Transformer model in the first 24 hours after the creation of the token.}
  \label{fig:metricsM2}
\ifadjust
\end{adjustwidth}
\fi
\end{figure}
  
  Figure \ref{fig:metricsM2} shows the evolution of the metrics for each of the ML algorithms used. XGBoost gets better metrics, except precision in some cases. We also notice that the metrics of the first hour are lower than those of the last. This confirms the intuition that our methods require a certain token history in order to work correctly and that models improve as this history grows. 
  Our algorithm obtains a very high accuracy (see \ref{section:ResultsSecondMethod}) even in the first hours. However, the precision, recall and f1-score are lower than in Activity based Method. In the best of cases, i.e. 20 hours after the creation of the pool, our best algorithm obtains a recall of 0.789. This could indicate that while malicious tokens are easily detectable in the first few hours, non-malicious tokens require more time. On the other hand, the precision remains quite high compared to the recall. This implies that, although the algorithms do not have a strong ability to detect non-malicious tokens, once they predict that one of them is non-malicious, it is very likely to be the case. 
 
\subsubsection{Unicrypt Results}
 As explained in Section \ref{Uniswap}, Unicrypt is a protocol that runs on top of the Uniswap protocol with the purpose of being a partial solution to rug pulls. In this work, we have empirically demonstrated that most of the tokens that use Unicrypt are malicious.
First, from our list of labelled tokens, 745 use Unicrypt, 725 are labelled as malicious and 20 as non-malicious.
Then, from the unlabelled tokens, we compute their features up to the present time and use the Activity based Method with XGBoost algorithm to evaluate them. Based on these predictions, 2544 of non-labelled tokens using Unicrypt, 2211 are predicted to be malicious and 333 non-malicious.

\section{Conclusions}\label{section:conclusions}
In summary, first, we increased the dataset provided in \cite{demystifying} by $18K$ scam tokens, finding new ways of actively executing the rug pulls. We then, we provided a theoretical classification to understand the different ways of executing the scam, and through the process of identifying rug pulls we found new token smart contract vulnerabilities (composability attacks) and new ways of money laundering. Based on this theoretical foundation, we provided a methodology to find rug pulls that had already been executed. Not surprisingly, we found that more than the $97,7\%$ of the tokens labelled were rug pulls. Finally, we defined two methods that use ML models to distinguish non-malicious tokens from malicious ones. We also verify the high effectiveness of these models in both scenarios. This implies that new malicious tokens can be detected prior to the malicious act,and, on the other hand, tokens supported by a strong project can also detected at an early stage. 

\section{Future Work}\label{section:Future_Work}
While our study has produced high precision and accuracy in detecting scams listed in Uniswap, it carries some limitations.
First, we believe that transferring learning techniques will not obtain the same quality results in DEXs of other chains such as PancakeSwap\footnote{https://pancakeswap.finance} and QuickSwap\footnote{https://quickswap.exchange}. Since the \texttt{gasPrice} is much lower, the economic cost of simulating volume and transfers is almost negligible. Therefore, in order to obtain similar results in other chains we should repeat the same longitudinal work and compute new features.  In addition, as market trends may change, these algorithms will have to be retrained in order to keep adding new information.\newline
Second, even though our approach for studying the source code of the tokens with Slither \cite{feist2019slither} was efficient and reliable for our purpose, it was not complete accurate, since it is a static analysis tool of the code and does not take into account complex composability problems among other protocols. Therefore, to have more insights for a particular token we suggest using stronger testing tools and formal verification tools, such as the ones  provided in \cite{babel2021clockwork}.\newline
Third, even though the clustering coefficient proved to be useful, computing this feature is highly time consuming. Therefore, we propose using other graph analysis, such as topology data analysis, to have a more efficient scam detection algorithm and obtain even more reliable insights into the transaction graph.\newline

\printbibliography
\appendix
\section{Table of features.}
\begin{table}[H]
\ifadjust
\begin{adjustwidth}{-5cm}{0cm}
\fi
\begin{tabular}{|
>{\columncolor[HTML]{C0C0C0}}c |
>{\columncolor[HTML]{EFEFEF}}c |c|}
\hline
Group                                                     & \cellcolor[HTML]{C0C0C0}Name & \cellcolor[HTML]{C0C0C0}Description                                                                \\ \hline
\cellcolor[HTML]{C0C0C0}                                  & liq\_curve                   & HHI applied to LP-tokens. \\ \cline{2-3} 
\multirow{-2}{*}{\cellcolor[HTML]{C0C0C0}HHI index}       & tx\_curve                    & HHI applied to each token.                                                 \\ \hline
\cellcolor[HTML]{C0C0C0}                                  & n\_pool\_syncs               & Total sync events.                                                                                 \\ \cline{2-3} 
\cellcolor[HTML]{C0C0C0}                                  & weth                         & Total weth.                                                                                        \\ \cline{2-3} 
\cellcolor[HTML]{C0C0C0}                                  & price                        & Price of token.                                                                                    \\ \cline{2-3} 
\multirow{-4}{*}{\cellcolor[HTML]{C0C0C0}Pool}            & liquidity                     & Total liquidity.                                                                                   \\ \hline
\cellcolor[HTML]{C0C0C0}                                  & lp\_transfer                 & Total number of LP-Token transfers.                                                                \\ \cline{2-3} 
\cellcolor[HTML]{C0C0C0}                                  & mints                        & Total number of mint events.                                                                       \\ \cline{2-3} 
\multirow{-3}{*}{\cellcolor[HTML]{C0C0C0}LP-Token}        & burns                        & Total number of burn events.                                                                       \\ \hline
\cellcolor[HTML]{C0C0C0}                                  & n\_transfers                 & Total number of transfers.                                                                         \\ \cline{2-3} 
\cellcolor[HTML]{C0C0C0}                                  & n\_unique\_addresses         & Total number of unique addresses.                                                                  \\ \cline{2-3} 
\multirow{-3}{*}{\cellcolor[HTML]{C0C0C0}Token transfers} & clus\_coeff                  & Clustering coefficient.                                                                            \\ \hline
\cellcolor[HTML]{C0C0C0}                                  & difference\_token\_pool     & Number of blocks between token and pool creation.                                   \\ \cline{2-3} 
\cellcolor[HTML]{C0C0C0}                                  & lock                         & This feature is 1 if part of the liquidity is locked and 0 otherwise.                                 \\ \cline{2-3} 
\cellcolor[HTML]{C0C0C0}                                  & yield                         & This features is 1 if there is yield farming involved and 0 otherwise.                                                                                                 \\ \cline{2-3} 
\multirow{-4}{*}{\cellcolor[HTML]{C0C0C0}Token}           & burn                         & This features is 1 if part of the liquidity has been burned and 0 otherwise.                                                                                                \\ \hline

\end{tabular}
\ifadjust
\end{adjustwidth}
\fi
\caption{This table describes each feature used in XGBoost and FT-Transformer classifiers. Note that apart from \textit{Token} group, all features are defined as time series.}
\end{table}
\section{Hyperparameters.} \label{section:Hyperparameters}
\begin{table}[H]
\centering
\begin{tabular}{|c|c|c|c|c|}
\hline
\rowcolor[HTML]{C0C0C0} 
Model & Parameter & Type & Distribution & Range \\ \hline
\cellcolor[HTML]{C0C0C0} & \cellcolor[HTML]{EFEFEF}max\_depth & Int & Uniform & {[}3, 10{]} \\ \cline{2-5} 
\cellcolor[HTML]{C0C0C0} & subsample & Float & Uniform & {[}0.5, 1{]} \\ \cline{2-5} 
\cellcolor[HTML]{C0C0C0} & learning\_rate & Float & Uniform & {[}1e-5, 1{]} \\ \cline{2-5} 
\cellcolor[HTML]{C0C0C0} & gamma & Float & Log-Uniform & {[}1e-8, 1e2{]} \\ \cline{2-5} 
\cellcolor[HTML]{C0C0C0} & lambda & Float & Log-Uniform & {[}1e-8, 1e2{]} \\ \cline{2-5} 
\multirow{-7}{*}{\cellcolor[HTML]{C0C0C0}XGBoost} & alpha & Float & Log-Uniform & {[}1e-8, 1e2{]} \\ \hline
\end{tabular}
\caption{List of hyperparameters optimized using the optuna Python library.}
\end{table}
\section{Second Method Results.}\label{section:ResultsSecondMethod}
\begin{table}[H]
\hspace{-0.1cm}
\begin{subtable}[c]{0.3\textwidth}
\begin{table}[H]
\ifadjust
\begin{adjustwidth}{-5cm}{0cm}
\fi
\begin{tabular}{ccccc}

\rowcolor[HTML]{C0C0C0} 
\multicolumn{5}{c}{\cellcolor[HTML]{C0C0C0}XGBoost 24h} \\
\rowcolor[HTML]{EFEFEF} 
Hour & Accuracy & Sensitivity & Precision & F1\_Score \\
1     &     0.990 &        0.714 &      0.810 &     0.758 \\
2     &     0.991 &        0.747 &      0.838 &     0.789 \\
3     &     0.991 &        0.755 &      0.837 &     0.793 \\
4     &     0.992 &        0.762 &      0.856 &     0.806 \\
5     &     0.992 &        0.773 &      0.849 &     0.809 \\
6     &     0.992 &        0.771 &      0.860 &     0.813 \\
7     &     0.992 &        0.763 &      0.851 &     0.804 \\
8     &     0.991 &        0.758 &      0.841 &     0.797 \\
9     &     0.991 &        0.762 &      0.845 &     0.801 \\
10    &     0.992 &        0.779 &      0.845 &     0.810 \\
11    &     0.992 &        0.773 &      0.849 &     0.808 \\
12    &     0.992 &        0.786 &      0.852 &     0.816 \\
13    &     0.992 &        0.778 &      0.859 &     0.815 \\
14    &     0.992 &        0.786 &      0.854 &     0.817 \\
15    &     0.992 &        0.776 &      0.851 &     0.811 \\
16    &     0.992 &        0.773 &      0.848 &     0.808 \\
17    &     0.992 &        0.779 &      0.852 &     0.813 \\
18    &     0.992 &        0.784 &      0.864 &     0.821 \\
19    &     0.992 &        0.782 &      0.858 &     0.818 \\
20    &     0.992 &        0.789 &      0.853 &     0.819 \\
21    &     0.992 &        0.779 &      0.863 &     0.818 \\
22    &     0.992 &        0.784 &      0.867 &     0.823 \\
23    &     0.992 &        0.784 &      0.869 &     0.824 \\
24    &     0.992 &        0.787 &      0.860 &     0.821 \\
\end{tabular}
\ifadjust
\end{adjustwidth}
\fi
\end{table}

\end{subtable}
\begin{subtable}[c]{0.3\textwidth}

\begin{table}[H]
\hspace{3.5cm}
\begin{tabular}{ccccc}
\rowcolor[HTML]{C0C0C0} 
\multicolumn{5}{c}{\cellcolor[HTML]{C0C0C0}FT-Transformer 24h} \\
\rowcolor[HTML]{EFEFEF} 
Hour & Accuracy & Sensitivity & Precision & F1\_Score \\
1     &     0.989 &        0.682 &      0.791 &     0.729 \\
2     &     0.988 &        0.702 &      0.778 &     0.734 \\
3     &     0.990 &        0.707 &      0.824 &     0.755 \\
4     &     0.987 &        0.707 &      0.751 &     0.724 \\
5     &     0.990 &        0.683 &      0.840 &     0.749 \\
6     &     0.989 &        0.706 &      0.805 &     0.749 \\
7     &     0.990 &        0.717 &      0.812 &     0.760 \\
8     &     0.990 &        0.710 &      0.821 &     0.759 \\
9     &     0.990 &        0.722 &      0.815 &     0.763 \\
10    &     0.990 &        0.712 &      0.828 &     0.762 \\
11    &     0.990 &        0.717 &      0.811 &     0.759 \\
12    &     0.989 &        0.722 &      0.795 &     0.755 \\
13    &     0.989 &        0.718 &      0.775 &     0.745 \\
14    &     0.989 &        0.746 &      0.789 &     0.764 \\
15    &     0.990 &        0.750 &      0.793 &     0.770 \\
16    &     0.990 &        0.734 &      0.812 &     0.767 \\
17    &     0.991 &        0.757 &      0.820 &     0.784 \\
18    &     0.991 &        0.746 &      0.826 &     0.782 \\
19    &     0.991 &        0.765 &      0.829 &     0.792 \\
20    &     0.990 &        0.720 &      0.838 &     0.765 \\
21    &     0.991 &        0.760 &      0.834 &     0.793 \\
22    &     0.991 &        0.747 &      0.834 &     0.786 \\
23    &     0.991 &        0.746 &      0.843 &     0.786 \\
24    &     0.991 &        0.758 &      0.819 &     0.785 \\
\end{tabular}
\end{table}
\end{subtable}
\end{table}
\end{document}